\documentstyle[12pt,oldlfont]{article}

\begin{document}

\newcommand{\non}{\nonumber}
\newcommand{\nn}{\noindent}
\newcommand{\s}{\\ \vspace*{-3mm}}
\newcommand{\ra}{\rightarrow}
\newcommand{\be}{\begin{eqnarray}}
\newcommand{\en}{\end{eqnarray}}

\renewcommand{\thefootnote}{\fnsymbol{footnote} }

\nn \hspace*{11cm} UdeM-GPP-TH-94-03 \\
    \hspace*{11cm} NYU--TH--94/06/03 \\
\hspace*{11cm} June 1994 \s

\vspace*{1.2cm}

\centerline{\large{\bf Leading electroweak correction
to Higgs boson}}

\vspace*{0.4cm}

\centerline{\large{\bf production at proton colliders.}}

\vspace*{1.5cm}

\centerline{\sc A.~Djouadi$^1$ and P.~Gambino$^2$.}

\vspace*{1cm}

\centerline{$^1$ Groupe de Physique des Particules,
Universit\'e de Montr\'eal,  Case 6128 Suc.~A,}
\centerline{H3C 3J7 Montr\'eal PQ, Canada.}

\vspace*{0.4cm}

\centerline{$^2$ Department of Physics, New York
University, 4 Washington Place,}
\centerline{ New York, NY 10003, USA.}

\vspace*{1.5cm}

\begin{center}
\parbox{13cm}
{\begin{center} ABSTRACT \end{center}
\vspace*{0.2cm}

\nn At proton colliders, Higgs particles are dominantly produced in the
gluon--gluon fusion mechanism. The Higgs--gluons coupling is mediated by heavy
quark loops, and the process can serve to count the number of heavy strongly
interacting particles whose masses are generated by the Higgs mechanism. We
present the two--loop leading electroweak radiative correction to this
coupling,
which is quadratically proportional to the heavy quark masses. It turns out
that
this correction is well under control across the physically interesting quark
mass ranges.}

\end{center}

\newpage

\renewcommand{\thefootnote}{\arabic{footnote}}

The fundamental particles, quarks, leptons and gauge bosons acquire their
masses through the Higgs mechanism \cite{Higgs}. This mechanism requires the
existence of at least one weak isodoublet scalar field, the self--interaction
of which leads to a non--zero field strength in the ground state, inducing the
spontaneous breaking of the SU(2)$\times$U(1) electroweak symmetry down to the
U(1) electromagnetic symmetry \cite{SM}. Among the four initial degrees of
freedom, three Goldstones will be absorbed to build up the longitudinal
polarization states of the massive $W^\pm$ and $Z$ bosons, and one degree of
freedom will be left over, corresponding to a physical scalar particle, the
Higgs boson. \s

The discovery of this particle is the most crucial test of the Standard Model
and the search for it will be one of the most important missions of future
high--energy colliders \cite{HHG,REV}. Unfortunately, in the Standard Model,
the Higgs boson mass $M_H$ is essentially a free parameter. The only
information available is the lower limit $M_H > 63.8$ GeV \cite{LEP}
established from the negative Higgs boson search in $Z$ boson decays at LEP;
this limit can be raised up to $\sim 80$ GeV in the second phase of LEP.
However, from the requirement of vacuum stability and from the assumption that
the Standard Model can be continued up to the Grand Unification scale, the
Higgs mass could well be expected \cite{GUT} in the window $80 < M_H < 180$
GeV, which is generally referred to as the intermediate mass range. \s

The dominant process for producing Higgs particles at proton colliders is the
gluon--gluon fusion mechanism \cite{gghborn}, $gg \rightarrow H$. The $Hgg$
amplitude is built up by heavy quark triangular loops, Fig.~1; in the minimal
Standard Model with three generations of fermions, the only significant
contribution is the one of the top quark. Since the quarks couple to the Higgs
bosons proportionally  to their masses, the loop particles will not decouple
from the amplitude when they are much heavier than the Higgs boson. This
coupling is therefore very interesting since it is sensitive to scales far
beyond the Higgs mass and can be used as a possible ``microscope" for new
strongly interacting particles whose masses are generated by the Higgs
mechanism. For instance, a fourth generation of fermions, the existence of
which is still allowed by present experimental data \cite{LEPDATA} if the
associated neutrino is heavy enough, would increase the $gg \ra H$ production
rate by an order of magnitude. \s

To lowest order, the $gg \ra H$ parton cross section can be expressed in
terms of a form factor derived from the quark triangle diagram in Fig.~1,
\begin{eqnarray}
\sigma^{\rm LO}(gg \ra H) = \frac{G_F \alpha_s^2}{288 \sqrt{2}\pi}
\left|\sum_Q F_Q (\tau_Q) \right|^2
\end{eqnarray}
with the form factor
\begin{eqnarray}
F_Q (\tau_Q)= \frac{3}{2}\tau_Q^{-1} \left[ 1+(1-\tau_Q^{-1}) \arcsin^2
\sqrt{\tau_Q} \right]
\end{eqnarray}

\nn approaching unity for quark masses slightly above half the Higgs boson
mass,
justifying the approximation of working in the limit $\tau_Q = M^2_H /4 m^2_Q
\ra 0$ already for $\tau_Q <1$. The Higgs boson production cross section for
proton colliders is found by integrating the parton cross section, eq.~(1),
over the gluons luminosity. \s

Because the precise knowledge of the $gg\rightarrow H$ production cross section
is mandatory, quantum corrections must be included. The QCD corrections have
been evaluated in Ref.~\cite{gghqcd} and found to be rather large,
increasing the production rate by more than 50\%. The next important radiative
correction to the $Hgg$ coupling, that is proportional to square of the masses
of the heavy quarks in the loop and is therefore potentially very large, is the
two--loop ${\cal O}(G_F m_Q^2)$ electroweak correction. In this Letter, we
present the result for this leading correction. We will work in the limit $m_Q
\ra \infty$ since, as mentioned previously, this is a very good approximation
for Higgs boson masses smaller than half the quark mass; this should hold at
least for Higgs bosons in the intermediate mass range\footnote{Note that up to
color and electroweak charges factors, the quark contribution to the $H Z
\gamma$ and $H \gamma \gamma$ couplings is the same as the one for the $Hgg$
coupling. At future $e^+ e^-$ colliders, the $H\gamma \gamma$ amplitude can be
precisely measured in the process $\gamma \gamma\rightarrow H$, the
high--energy photons being generated by Compton--back scattering of laser light
\cite{gamma}; this amplitude is also important since the $\gamma \gamma$ decay
of the Higgs boson is the most promising detection channel of this particle at
hadron colliders. The leading ${\cal O}(G_F m_Q^2)$ correction to the $Hgg$
amplitude presented here, will be the same for the $H\gamma \gamma$ and $HZ
\gamma$ amplitudes.}. \s

The technique that we use to calculate the two--loop ${\cal O}(G_F m_Q^2)$
correction to the $Hgg$ coupling has been known for some time \cite{LET,ACD}.
Writing the basic Higgs--quark Lagrangian as
\be
{\cal L} (HQ \bar{Q}) = -(\sqrt{2}G_F)^{1/2} \, m_Q^0 \, H Q_0 \bar{Q}_0
\en
the $Hgg$ coupling at small momentum can be derived from the condition that
the matrix element, $\langle gg| \theta_\mu^\mu|0 \rangle$, of the trace of
the energy--momentum tensor
\be
\theta_\mu^\mu = (1- \delta_2) \, m_Q^0 \, Q \bar{Q}+ \frac{1}{2} \, \frac{
\beta  (\alpha_S)}{g_S} \, G_{\mu \nu} G^{\mu \nu}
\en

\vspace*{2mm}

\nn vanishes in the low--energy limit. Here, $G_{\mu \nu}$ is the gluon field
strength tensor, $\alpha_S=g_S^2/4\pi$ with $g_S$ the strong coupling constant
and $\beta (\alpha_S)$ is the QCD $\beta$ function to which a quark
contributes by an amount
\be
\frac{ \beta (\alpha_S)}{g_S} = \frac{\alpha_S}{6\pi} \, \left[ \, 1 +
\delta_1 \, \right ]
\en
where the term $\delta_1$ denotes the higher--order contribution. To evaluate
this contribution  at ${\cal O}(\alpha_S G_Fm_Q^2)$, one needs to consider the
two--loop diagrams shown in Fig.~2 and the corresponding counterterms. In
renormalizable gauges, the virtual scalar bosons exchanged in the loops
correspond to either the Higgs boson or to the neutral and charged Goldstone
bosons. Note that in the amplitude for a quark of a given flavor, the virtual
exchange of the charged Goldstone boson will introduce the weak isospin partner
of this quark.

The term $\delta_2$ in eq.~(4) arises from a subtlety in the use of the
low--energy theorem \cite{ACD}: in renormalizing the $H Q \bar{Q}$ interaction,
eq.~(3), the counterterm for the Higgs--quark Yukawa coupling is not the
$HQ\bar{Q}$ vertex with a subtraction at zero momentum transfer, $\Gamma_{HQ
\bar{Q}}(q^2=0)$ [which is implicitly used in the low--energy theorem], but
rather is determined by the counterterms for the quark mass $\delta m_Q$ and
quark wave--function $Z_2^Q$. This has to be corrected for, and one then has
\be
\delta_2 = (Z_2^Q -1) - \frac{\delta m_Q}{m_Q}  + \Gamma_{H Q \bar{Q}} (q^2=0)
\en

Finally, one needs to include the renormalization of the Higgs boson wave
function and the vacuum expectation value of the Higgs field. This is achieved
by multiplying the one--loop generated $Hgg$ coupling by a factor $1+\delta_3$
where, in terms of the $W$ and $H$ boson vacuum polarization functions at
zero--momentum transfer, $\delta_3$ reads
\be
\delta_3= - \frac{1}{2} \left[ \frac{ \Pi_{WW}(0)}{M_W^2} + \left. \frac{
\partial \Pi_{HH} (M_H^2)}{\partial M_H^2} \right|_{M_H^2 \ra 0} \  \right]
\en
The complete ${\cal O}(G_Fm_Q^2)$ correction to the effective $Hgg$ coupling
will be then given by
\be
{\cal L}(Hgg)= (\sqrt{2}G_F )^{1/2} \, \frac{\alpha_S}{12\pi} \,
\, HG_{\mu \nu} G^{\mu \nu} \, (1+\delta)
\en
with

\vspace*{-1.2cm}

\be
\delta= \delta_1 + \delta_2 +\delta_3
\en
and the corrected $gg \ra H$ cross section at this order will read
\be
\sigma (gg \ra H) = \sigma^{\rm LO} (gg \ra H) \, [ \, 1+ 2\delta \, ]
\en

Using dimensional regularization, we have evaluated the contribution of the
diagrams shown in Fig.~2 as well as those of the various one--loop self--energy
and vertex functions which enter the counterterms in $\delta_1$ and the terms
$\delta_2$ and $\delta_3$, in the case of a weak isodoublet of heavy quarks
$(U,D)$ with masses $m_U \neq m_D$. The calculation has been performed in the
on--shell scheme which is usually used in the electroweak theory \cite{sir};
in this scheme, the quark masses correspond to the poles of the quark
propagators. \s

We have then specialized to two particular cases of physical relevance: $(i)$
$m_U \gg m_D$ which corresponds to the approximate contribution of the
top--bottom weak isodoublet since $m_t \sim 174$ GeV \cite{CDFTOP} is much
larger than $m_b \sim 5$ GeV and $(ii)$ $m_U \sim m_D$ which corresponds to the
contribution of an additional generation of fermions since in this case, the
mass splitting between the members of the extra weak isodoublet is highly
constrained by electroweak precision measurements \cite{LEPDATA}. The lengthy
results in the general case $m_U \neq m_D$ as well as the tedious details of
the calculation will be given elsewhere \cite{nous}; in this short Letter we
will simply present our final results in the two special cases of interest.

In the minimal Standard Model with three fermion families, the ${\cal O}(
\alpha_S G_F m_t^2)$ contribution to the top quark loop amplitude in the limit
$m_t \gg m_b$ is given by
\be
\delta = + \frac{G_F \sqrt{2}}{32 \pi^2} m_t^2
\en
Due to a large cancellation among the various $\delta_i$ contributions [in
units of $ \delta /m_t^2$ one has: $\delta_1 = -12, \ \delta_2=+ 6$ and
$\delta_3=7$], the total correction is very small: for a value $m_t \sim 200$
GeV, which can be viewed as a conservative upper bound on the top quark mass
\cite{CDFTOP}, it amounts to a mere [positive contribution of] 0.2\%.
Therefore, contrary to the QCD corrections which have been found to be very
large \cite{gghqcd}, the leading electroweak correction to the top quark loop
mediated Higgs--gluons coupling turns out to be very small.
Note that the correction is free of infrared singularities for $m_b \ra 0$, as
required by the Kinoshita--Lee--Nauenberg theorem \cite{KLN}. \s

In the case of a fourth family of heavy quarks with degenerate masses, $m_U =
m_D =m_Q$, the ${\cal O}(G_F m_Q^2)$ correction to one of the quarks amplitude
is given by\footnote{The calculation in the equal mass case has been first
performed in Ref.~\cite{Hoogeveen}. However, only the irreducible contribution
$\delta_1$ [including quark mass, wave--function and vertex renormalizations
with a subtraction at zero--momentum transfer for the Higgs--quarks vertex] has
been evaluated: the proper renormalization of the Higgs--quarks Yukawa coupling
and the renormalizations of the Higgs wave--function and vacuum expectation
value have been omitted. As a consequence, the result of Ref.~\cite{Hoogeveen}
is a factor of three larger compared to our result.\s }
\be
\delta = - \frac{G_F \sqrt{2}}{8 \pi^2} m_Q^2
\en

This negative correction will therefore screen the value of the one--loop
generated $Hgg$ coupling. However, the correction is rather small since for
realistic values of the quark masses\footnote{Since the new fermions aquire
their masses through the standard Higgs mechanism, upper bounds on the masses
can be derived from imposing partial wave unitarity on their scattering
amplitudes: in the production of longitudinal $W/Z$ or $H$ bosons in $F\bar{F}$
scattering at high--energies, weak interactions become strong and perturbation
theory breaks down if $m_F$ becomes too large. In the tree level approximation,
an upper bound of $m_Q <500$ GeV can be obtained for a fourth generation quarks
\cite{CFH}. \s}, $m_Q < 500$ GeV, it does not exceed the 5\% level. It is only
for quark masses larger than $\sim 2$ TeV, for which perturbation theory breaks
down already at the tree level \cite{CFH}, that the radiative correction will
exceed the one--loop result. Therefore, the ${\cal O}(G_F m_Q^2)$ correction to
the $Hgg$ amplitude is well under control for quark masses in the range
interesting for perturbation theory, and the counting of new heavy quarks via
the $Hgg$ coupling will not be jeopardized by these radiative corrections. \s

Note that in the previous equation only the contribution of the heavy quarks of
the fourth generation has been taken into account. Additional contributions
will be induced by the extra weak isodoublet of leptons [with a right--handed
component for the heavy neutrino, for the mass of the latter particle to be
generated through the standard Higgs mechanism] via the renormalization of the
Higgs boson wave--function and the one of the vacuum expectation value of
the Higgs field. If one assumes that the masses of the heavy leptons are
approximately equal to those of the quarks, the total contribution of the
weak isodoublets of quarks and leptons to the coefficient $\delta$ will be
smaller by a factor of three than in eq.~(12). \s

Finally, we observe that in this equal mass case, the quark mass
renormalization
does not contribute to the amplitude in the limit $m_Q\rightarrow \infty$, and
therefore the result for the correction $\delta$ is independent on the scheme
in which the quark mass is defined. This can be understood by recalling that in
this limit, the quark contribution to the one--loop amplitude decouples in the
sense that there is no more dependence on the quark mass. \s

In conclusion. We have presented the two--loop leading ${\cal O}(G_F m_Q^2)$
electroweak radiative correction to the Higgs--gluon--gluon coupling. This
coupling is very interesting since it is sensitive to scales far beyond the
Higgs boson mass. In the case of the minimal Standard Model with only three
fermion families, the correction to the heavy top quark contribution is very
small: less than 0.2\% for a top quark mass smaller than 200 GeV. If the
Standard Model is extended to include a fourth generation of heavy fermions,
the corrections to the additional quark loop amplitudes are well under control
across the physically interesting quark mass ranges for perturbation theory,
since in this case they do not exceed the 5\% level.

\vspace*{0.5cm}

\nn {\bf Acknowledgements} \s

\nn  Discussions with G. Degrassi, S. Fanchiotti, A. Sirlin, M. Spira and P.
M. Zerwas are gratefully acknowledged. This work is supported by the Natural
Sciences and Engineering Research Council of Canada and by the National Science
foundation under Grant No. PHY--9313781.

\end{document}